\documentclass[preprint,showpacs,preprintnumbers,amsmath,amssymb]{revtex4}

\usepackage{graphicx}% Include figure files
\usepackage{dcolumn}% Align table columns on decimal point
\usepackage{bm}% bold math
\usepackage{epsfig}
\newcommand{\be}{\begin{eqnarray}}
\newcommand{\ee}{\end{eqnarray}}

\def\i{\int_{-\infty}^{\infty}}

\begin{document}

\title{Universal  Tomonaga-Luttinger liquid phases in
one-dimensional strongly attractive $SU(N)$  fermionic cold
atoms}

\author{ X.W. Guan$^{1}$, J.-Y. Lee$^{1}$, M.T. Batchelor$^{1,2}$,
 X.-G. Yin$^{3}$ and  S. Chen$^{3}$ 
}
\affiliation{${1}$ Department of Theoretical Physics, Research
 School of Physics and Engineering,
Australian National University, Canberra ACT 0200,  Australia}
\affiliation{${2}$ Mathematical Sciences Institute,
Australian National University, Canberra ACT 0200,  Australia}
\affiliation{${3}$ Institute of Physics, Chinese Academy of Sciences,
  Beijing 100190, China}

\date{\today}

\begin{abstract}
A simple set of algebraic equations is derived for the 
exact low-temperature thermodynamics of one-dimensional
multi-component strongly attractive fermionic atoms with
enlarged $SU(N)$ spin symmetry and Zeeman splitting.
Universal multi-component Tomonaga-Luttinger liquid
(TLL) phases are thus determined.
For linear Zeeman splitting, the physics of the gapless phase at low
temperatures belongs to the  universality class of a two-component
asymmetric TLL corresponding to spin-neutral $N$-atom
composites and spin-$(N-1)/2$ single atoms.
The equation of states is also obtained to open up the study of 
multi-component TLL phases in 1D systems of $N$-component 
Fermi gases with population imbalance.
\end{abstract}



\maketitle

A significant feature of one-dimensional (1D) many-body systems is the
universal low energy physics of the gapless phase described by a 
Tomonaga-Luttinger liquid (TLL) \cite{Giamarchi-b}, as recently revealed in 
experimental measurements on the thermodynamics of spin ladder materials
\cite{LL-exp}. In such systems magnetic fields drive phase transitions 
between gapped and gapless phases.
In the gapless phase spin excitations (spinons) carrying spin-$1/2$
give rise  to universal TLL thermodynamics.

On the other hand, exquisite control over the effective spin-spin
interaction between cold atoms \cite{Li} has provided a new
opportunity to rigorously test spin liquid behaviour via trapped
fermionic atoms with higher spin symmetry %in a clean environment
\cite{Ho,Rapp,Lecheminant,quatet}.
In particular, fermionic alkaline-earth atoms display an exact
$SU(N)$ spin symmetry with $N=2I+1$ where $I$ is the nuclear
spin \cite{Gorshkov}.
For example, $I=9/2$ for ${}^{87}$Sr and $I=5/2$ for ${}^{171}$Yb.
Such fermionic systems with enlarged $SU(N)$ spin symmetry 
are expected to display a remarkable diversity of new quantum phases 
%and phase transitions 
due to the existence of multiple charge bound states.

We derive the universal thermodynamics of
the gapless phase in strongly attractive 1D fermions  
with $N$-component hyperfine states.
We thus find that population imbalances controlled by Zeeman
splitting can be used to explore and control multi-component TLL physics 
 in 1D interacting fermionic gases with $SU(N)$ spin symmetry.
The universal crossover from the regime of a relativistic
multi-component Luttinger liquid to a nonrelativistic quantum critical
regime is determined  from the specific heat phase diagrams.

In principle, the thermodynamic Bethe ansatz (TBA) can provide the exact
thermodynamics of such systems.
However, the TBA involves an infinite number of coupled nonlinear
integral equations which hinders access to the thermodynamics 
from both the numerical and analytical points of view
\cite{kakashvili,casula}.
To  overcome this obstacle, a method  was proposed to analytically
obtain thermodynamic quantities of the  two-component attractive
Fermi gas \cite{ZGLO}.
We build on this approach to reduce the multiple charge bound state TBA for  
attractive 1D fermions with arbitrary $SU(N)$ spin symmetry 
 to a set of simple algebraic equations to provide universal
and exact low-temperature thermodynamics of the model.

{\it The model.} We consider a system of
$N_f$ interacting fermions of equal mass $m$ which may occupy $N$
possible hyperfine levels ($|i\rangle$, $i=1,\ldots,N$) labeled by
$N$ isospin states
and constrained by periodic boundary conditions to a line of length $L$.
The Hamiltonian \cite{Sutherland,Takahashi} is
\begin{equation}
{\cal{H}}=-\frac{\hbar ^2}{2m}\sum_{i = 1}^{N_f}\frac{\partial
^2}{\partial x_i^2}+g_{\rm 1D} \sum_{1\leq i<j\leq N_f} \delta
(x_i-x_j)+E_z \label{Ham}
\end{equation}
with Zeeman energy
$E_z=\sum_{i=1}^{N}N^{i}_f\epsilon^{i}_Z(\mu_B^{i},B)$.
Here $N^{i}_f$ is the number of fermions in state $| i\rangle$
with Zeeman energy $\epsilon^{i}_Z$ (or say, a chemical potential
$\mu_i$) determined by the magnetic
moments $\mu_B^{i}$ and the magnetic field $B$.
The spin-independent contact interaction $g_{\rm 1D}$ remains between
fermions with different hyperfine states so that the spins are
conserved, i.e., $N^i_f$ with $i=1,\ldots,N$ are good quantum
numbers.
The coupling constant $g_{\rm 1D} ={\hbar ^2 c}/{m}$ with
interaction strength $c=-{2}/{a_{\rm 1D}}$ determined by the
effective 1D scattering length $a_{\rm 1D}$.
For simplicity, we choose the dimensionless units of $\hbar = 2m = 1$
for numerical calculation and use the dimensionless coupling constant $\gamma=c/n$ with linear
density $n ={N_f}/{L}$.
The Hamiltonian (\ref{Ham}) exhibits $U(1)\times SU(N)$ symmetry
where $U(1)$ is associated with the charge degree of freedom and
$SU(N)$  with the $N$ hyperfine spin states.

\begin{figure}[t]
{{\includegraphics [width=0.8\linewidth]{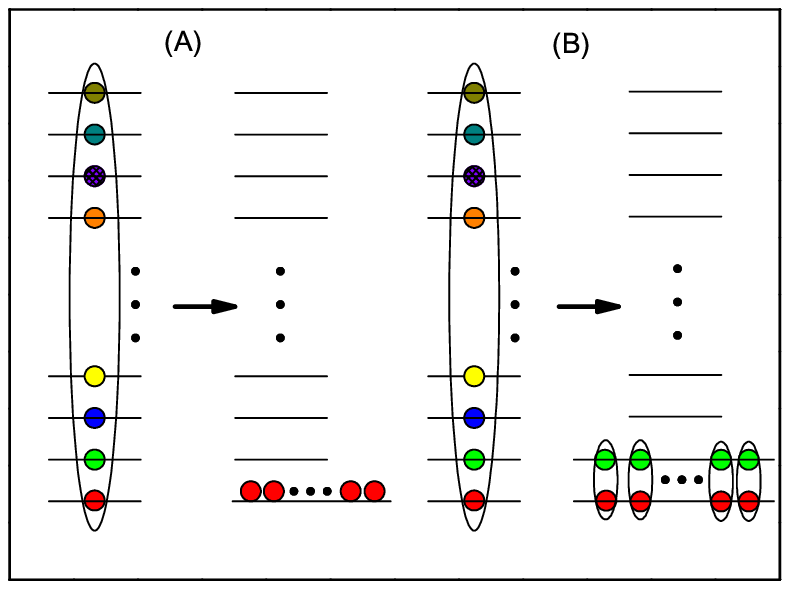}}} 
\caption{Phase
transitions from the bound states of $N$-fermions
  into (A) normal Fermi liquid and (B) paired states are induced by linear
  and nonlinear Zeeman splittings,
  respectively. Ellipses denote the charge  bound
  states.}\label{fig:states}
\end{figure}

{\it Ground states.}
The energy eigenspectrum is given in
terms of the fermion quasimomenta $\left\{k_i\right\}$ 
satisfying the Bethe ansatz equations (BAE) \cite{Sutherland,Takahashi} 
by $E=\sum_{j=1}^{N_f} k_j^2$.
For attractive interaction, the BAE  allow charge bound states and spin strings.
The $SU(N)$ symmetry acquires $N-1$ kinds of charge bound
states which can be viewed as composites of $r$-atoms with total
spin $s$, where $s=r(N-r)/2$ and $r=2,\ldots,N$.
In order to simplify calculations in the study of magnetism, we
rewrite the Zeeman energy as $E_Z=-\sum_{i=1}^{N-1}H_iN_i$
where the independent parameters $H_i$ with $i=1,\ldots, N-1$
characterize the  chemical potentials for $N_1$ unpaired fermions
and  $N_i$  charge bound states of $i$-atoms.
Here we set up explicit relations among the $H_i$'s and arbitrary (nonlinear) Zeeman splittings
$\Delta_{i+1\,i}=\epsilon^{i+1}_Z-\epsilon^{i}_Z$ via the relations
$\Delta_{i+1\,i}=-H_{i-1}+2H_i-H_{i+1}$
with $H_{N+1}=0$.
We shall show that equally spaced (linear) Zeeman splitting,
i.e., $\Delta_{a+1\,a}=H$ for $a=1,\ldots,N-1$, drives a
smooth phase transition from bound states of $N$-atoms into a
normal Fermi liquid at zero temperature, see Fig.~\ref{fig:states}(A). 
Nonlinear Zeeman splittings may trigger spin-$s$ charge bound states 
as illustrated in Fig.~\ref{fig:states}(B), where the magnetic moments of paired states is
$s=N-2$ \cite{Ho}.

For strong attraction ($|\gamma| \gg 1$) these charge bound states
are stable and the system is strongly correlated.
The corresponding binding energies of the charge bound states are
 given by $\epsilon_{r} = {\hbar^2c^2r(r^2-1)}/ {(24m)}$.
Solving the BAE with strongly attractive interaction, 
the ground state energy $E_0^{\infty}$ per unit length 
in the thermodynamic limit is given explicitly by (in units of $\hbar^2/2m$)
\begin{equation}
 E_0^{\infty} \approx
\sum_{r=1}^{N}\frac{\pi^2n_r^3}{3r}\left(1+\frac{2}{|c|}A_r+\frac{3}{c^2}A_r^2\right)-\sum_{r=2}^{N}n_{r}
   \epsilon_{r}.\label{E}
\end{equation}
Here  $n_r=N_r/L$ and we have defined the functions
\begin{equation}
A_{r}=\sum_{j=1}^{r-1}\sum_{i=j}^{N}\frac{4n_{i}\theta(r-2)}{r(i+r-2j)}+
\sum_{i=r+1}^{N}\frac{4n_{i}\theta(N-r-1)}{r(i-r)}, 
\end{equation}
where $\theta(x)$ is the step function.

The general result (\ref{E}) for arbitrary $N$ unifies and extends known results  \cite{GB,Wadati,Hu} 
for isospin $S=1/2, 1$ and $3/2$.
Furthermore, we find that the low-lying excitations are described by the 
linear dispersion relations $\epsilon^{r}(k)=v_r(k-k_F^r)$, where
$v_r\approx \frac{\hbar \pi
n_r}{mr}(1+\frac{2}{|c|}A_r+\frac{3}{c^2}A_r^2)$ with $r=1,\ldots,
N$ are the velocities for unpaired fermions and charge bound
states of $r$-atoms. We denote the corresponding Fermi momentum
by $k_F^r$.  These dispersion relations lead naturally to the universal form \cite{Cardy,Affleck} for 
finite-size corrections to the ground state energy
\begin{equation}
E(L,N_f)-LE_0^{\infty} \approx -\frac{\pi \hbar
C}{6L}\sum_{r=1}^{N}v_r.\label{FSC}
\end{equation}
The central charge $C=1$ for $U(1)$ symmetry.  Here the universal
 finite-size corrections (\ref{FSC}) indicate
 the TLL signature of 1D many-body physics \cite{Giamarchi-b}.

{\it Quantum phase transitions.} 
At finite temperatures, 
the equilibrium states are determined by the minimization of the Gibbs
free energy \cite{Takahashi-B}, which gives rise to a set of
coupled nonlinear integral equations -- the TBA equations \cite{note}.
The TBA equations provide not only finite temperature thermodynamics
but also the band fillings with respect to Zeeman splittings  and
chemical potentials which may be conveniently used to analyze  quantum
phase transitions at zero temperature.
From the TBA equations, we find that complete phase diagrams at zero
temperature are determined by the independent external
field-energy transfer relations
\begin{equation}
H_r=\frac{1}{12}{r(N^2-r^2)c^2}+r(\mu^{(r)}-\mu^{N}),\label{Hr}
\end{equation}
for $r=1,\ldots,N-1$.
The effective chemical potentials $\mu^{(r)}
=\mu +\frac{H_r}{r}+\frac{(r^2-1)}{3}\frac{c^2}{4}$ with
$r=1,\ldots,N$ are given by
\begin{equation}
\frac{\mu^{
    (r)}}{\pi^2}\approx\frac{n_{r}^2}{r^2}\left(1+\frac{2A_{r}}{|c|}+\frac{3A_{r}^2}{c^2}\right)+\frac{\vec{B}_{
    r} \cdot \vec{I}}{|c|}+\frac{3\vec{B}_{r} \cdot \vec{A}}{c^2},\label{mu}
\end{equation}
characterizing Fermi surfaces of stable charge bound states and
unpaired fermions.
Here $\vec{A}=(A_1,\ldots,A_{N})$ and
$\vec{B}_{r}=(B_{r}^1,\ldots,B_{r}^{N})$ with 
\begin{equation}
B_{r}^{\ell}
=\frac{8n_{\ell}^3}{3r \ell}\left[\sum_{j=1}^{\ell-1}
\frac{\theta(r-j)}{\ell(r+\ell-2j)}+\frac{\theta(r-\ell-1)}{\ell
(r-\ell)}\right]
\end{equation}
for $r,\ell=1,\ldots,N$. $\vec{I}$ is a unit vector.
The energy transfer relations (\ref{Hr}) can be used to show 
that linear Zeeman splitting  can  only lift $SU(N)$ symmetry to
$U(1)^{N}$ symmetry, recall Fig.~\ref{fig:states}.

\begin{figure}[t]
{{\includegraphics [width=1.0\linewidth]{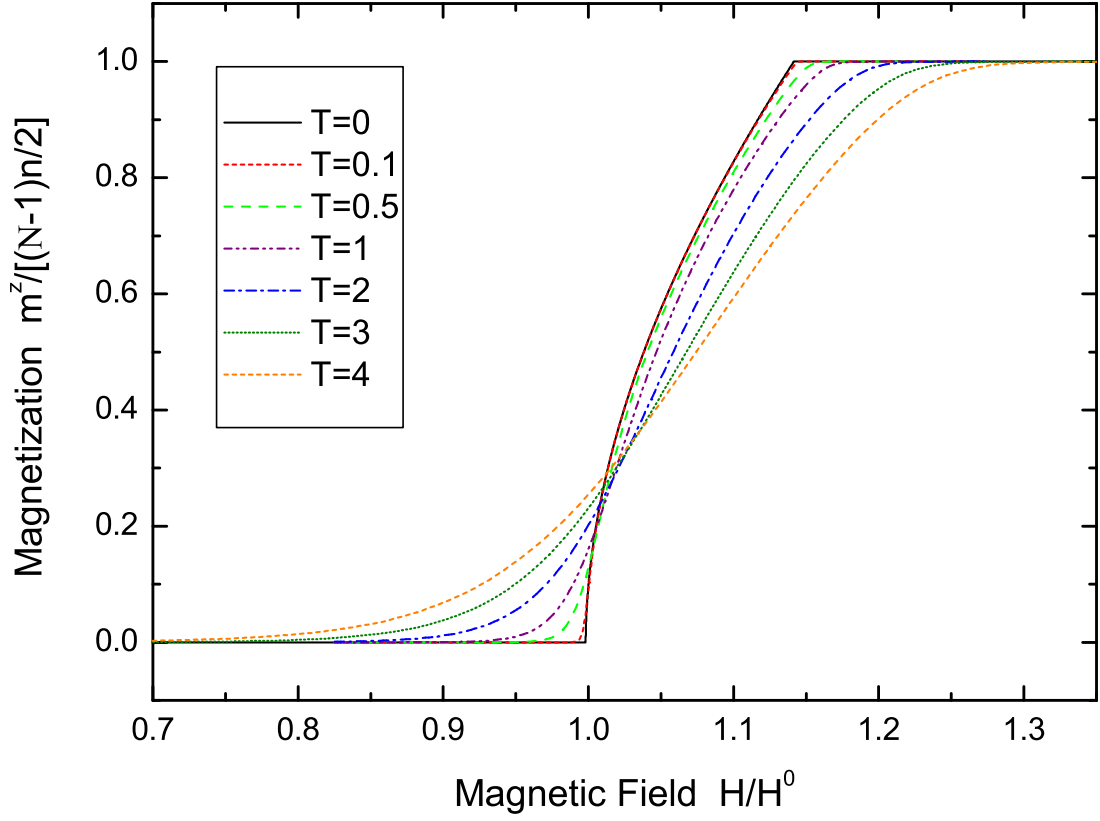}}}
\caption{Typical magnetization vs external field for $SU(3)$ 
  fermions with strong coupling ($c=-10$ and $n=1$) at
  different temperatures. The curves for $SU(N)$ are similar.
  }\label{fig:mz}
\end{figure}

For pure Zeeman splitting, if the external field $H$ is less than a
lower critical field $H_{c1}$, a molecular superfluid phase of spin-neutral
bound states forms the ground state.  We find the value
\begin{eqnarray}
H_{c1}&\approx&\frac{(N+1)n^2\gamma^{2}}{6}-\frac{2\pi^{2}n^{2}}{N^{4}(N-1)}\left[1+\frac{16Q}{3N^{2}|\gamma|}\right.\nonumber\\
& &\left.-\frac{8}{3N(N-1)|\gamma|}+\frac{20Q^2}{\gamma^2N^4}- \frac{16Q}{\gamma^2N^3(N-1)}\right]
\end{eqnarray}
at which the spin gap is diminished. %, thus the excitation becomes gapless. 
Here $Q=\sum_{j=1}^{N-1}\frac{1}{N-j}$.
On the other hand, when $H > H_{c2}$, where
\begin{equation}
H_{c2}\approx \frac{(N+1)n^{2}\gamma^{2}}{6}+\frac{2\pi^{2}n^{2}}{(N-1)}\left[1-\frac{8}{3N(N-1)|\gamma|}\right]
\end{equation}
 the system is fully-polarized into a normal Fermi liquid.  
For the 
intermediate regime $H_{c1}< H <H_{c2}$ spin-neutral bound
states of $N$-atoms and unpaired fermions coexist.
The magnetization gradually increases from $m^z=0$ to its
normalized value $m^z=1$ as the field increases from $H_{c1}$ to $H_{c2}$ 
(see the solid line Fig.~\ref{fig:mz}).  More subtle phases may be explored 
using the energy-transfer relations (\ref{Hr}) by controlling the 
nonlinear Zeeman splitting parameters, e.g., the BCS pairing
phase (see Fig.~\ref{fig:states}) by setting $\Delta_{2\,1}=h_1$  and
$\Delta_{a+1\,a}=h_2$ for $a=2,N-1$.  
All phase transitions are of second order with a
linear field-dependent magnetization in the vicinity of critical
points due to the condition of fixed total particle number.

{\it Universal thermodynamics.}
At high temperatures $T\sim \epsilon_r$ (setting the Boltzmann constant to unity), 
thermal fluctuations  can break the charge bound states while  spin
fluctuations lead to an effective ferromagnetic spin-spin interaction
coupled to each  Fermi sea of spin-$s$ charge bound states and unpaired
fermions.
The effective ferromagnetic spin-spin coupling constants  are  given by
$J^{(r)}=\frac{2}{r|c|}p^{(r)}$ for $r=1,\,2,\ldots,N-1$ \cite{note}.
In this sense, we may simply view  the non-neutral charge bound
state as a molecule  with spin $s$, which could flip its spin
antiparallel to the external fields $H_r$ due to thermal fluctuations.
However, such spin fluctuations  coupled to
the channels of unpaired fermions and the spin-$s$ charge bound
states are suppressed by large fields $H_r$ at low temperatures.
Thus the low energy physics is dominated by charge density fluctuations.
Indeed we show that in the physically interesting regime
$T\ll \epsilon_r, \, \Delta_{i+1\,i},\,\, \gamma\gg 1$
 the breaking of charge  bound states
and spin wave fluctuations  is strongly suppressed.
Thus each dressed energy can be written in a single particle form
$\epsilon^r(k)=\hbar^2rk^2/2m-\bar{\mu}^{(r)}+O(1/\gamma^3)$,
where the marginal scattering energies among composites and unpaired
fermions and spin-wave thermal fluctuations are considered in the
chemical potentials $\bar{\mu}^{(r)}$.

The thermodynamics at finite temperatures  can thus be obtained from the
set of algebraic equations \cite{note}
\begin{eqnarray}
\bar{\mu}^{(r)}&\approx &r\mu^{(r)}-\sum^r_{j=1}\sum^{N}_{\scriptscriptstyle{\small
    \begin{array}{c}i=j\\i\ne
2j-i\end{array}}}\frac{4P^{(i)}}{i(i+r-2j)|c|}+f_s^{(r)},\nonumber \\
p^{(r)}&\approx
&-\sqrt{\frac{rm}{2\pi\hbar^2}} \, T^{\frac{3}{2}} \, {\rm Li}_{\frac{3}{2}}\left(-e^{\bar{\mu}^{(r)}/T}\right)
 \label{TBA}
\end{eqnarray}
for $r=1,\ldots,N$.
Here ${\rm Li}_{s}(x)$ is the standard polylogarithm function and
$f_s^{(r)}$ denotes the free energy of an effective ferromagnetic
spin-spin interaction with coupling constant $J^{(r)}$ in the channel
of the spin-$s$ charge bound state \cite{note}. However, because
it is a spin singlet, there is no such
effective spin-spin interaction for the spin-neutral bound state of
$N$-atoms, i.e., $f_s^{(N)}=0$ and $H_{N}=0$.  
$p^{(r)}$ is the pressure for charge bound states of $r$-atoms.
The pressure for the system is given by $p=\sum_{r=1}^{N}p^{(r)}$.
The spin string contributions to thermal
fluctuations in the physically interesting  regime can be asymptotically calculated from the spin string
equations for these non-spin-neutral charge bound states and unpaired
fermions, i.e., $f_s^{(r)}\approx
Te^{-\frac{\Delta_{r+1\,r}}{T}}e^{-\frac{J^{(r)}}{T}}I_0(\frac{J^{(r)}}{T})$, where 
$I_0(x)=\sum_{k=0}^{\infty}\frac{1}{(k!)^2}(\frac{x}{2})^{2k}$.
It is important to observe that 
$f_s^{(r)}$ becomes exponentially small as $T \to 0$.

The suppression of spin fluctuations leads to
a universality class of a multi-component TLL  in each gapless
phase, where the charge bound states of $r$-atoms form hard-core
composite particles.
In order to see this  universal TLL physics,  we calculate
the leading low temperature corrections to the free energy by using
Sommerfeld expansion with the TBA equations to give
\begin{equation}
f \approx f_0-\frac{\pi
  T^2}{6\hbar}\sum_{r=1}^{N}\frac{1}{v_{r}}. \label{FreeE}
\end{equation}
This result  is consistent
with the finite-size correction result  (\ref{FSC}). In the above
equation $f_0= E_0^{\infty}-\sum_{r=1}^{N-1}n_rH_r$.
This result proves the existence of TLL phases in 1D gapped systems at
low temperatures. Although there is no quantum phase transition in 1D
many-body systems at finite temperatures due to thermal 
fluctuations, the existence of the TLL leads to a crossover
from relativistic dispersion to nonrelativistic dispersion between
different regimes at low temperatures \cite{Maeda}.
Nevertheless, we find that such a field-induced multi-component
TLL  only lies in a small portion of Zeeman parameter space.
Linear Zeeman splitting may result in a two-component Luttinger
liquid in a large portion of Zeeman parameter space at low temperatures.

The thermodynamics of the gapless phase for the model (\ref{Ham}) 
can be analytically calculated with
linear Zeeman splitting, i.e. with  $\Delta_{a+1\,a}=H$ for
$a=1,\ldots,N-1$. In the regime
$H_{c1}<H<H_{c2}$  the TBA equations (\ref{TBA}) reduce to two coupled
equations for $ \bar{\mu}^{(1)}$ and $\bar{\mu}^{(N)}$. The rest
of the effective  $\bar{\mu}^{(r)}=0$.
From these two equations, we can obtain the density $n=\partial
p/\partial \mu$, magnetization $m^z=\partial p/\partial H$ and
free energy $f=\mu n-p$ by iteration. Fig.~\ref{fig:mz} shows 
the magnetization for attractive three-component fermions 
with a pure Zeeman field at different temperatures.
We see that the linear field-dependent phase transitions in the
vicinity of the critical points are smeared. 

\begin{figure}[t]
{{\includegraphics [width=1.0\linewidth]{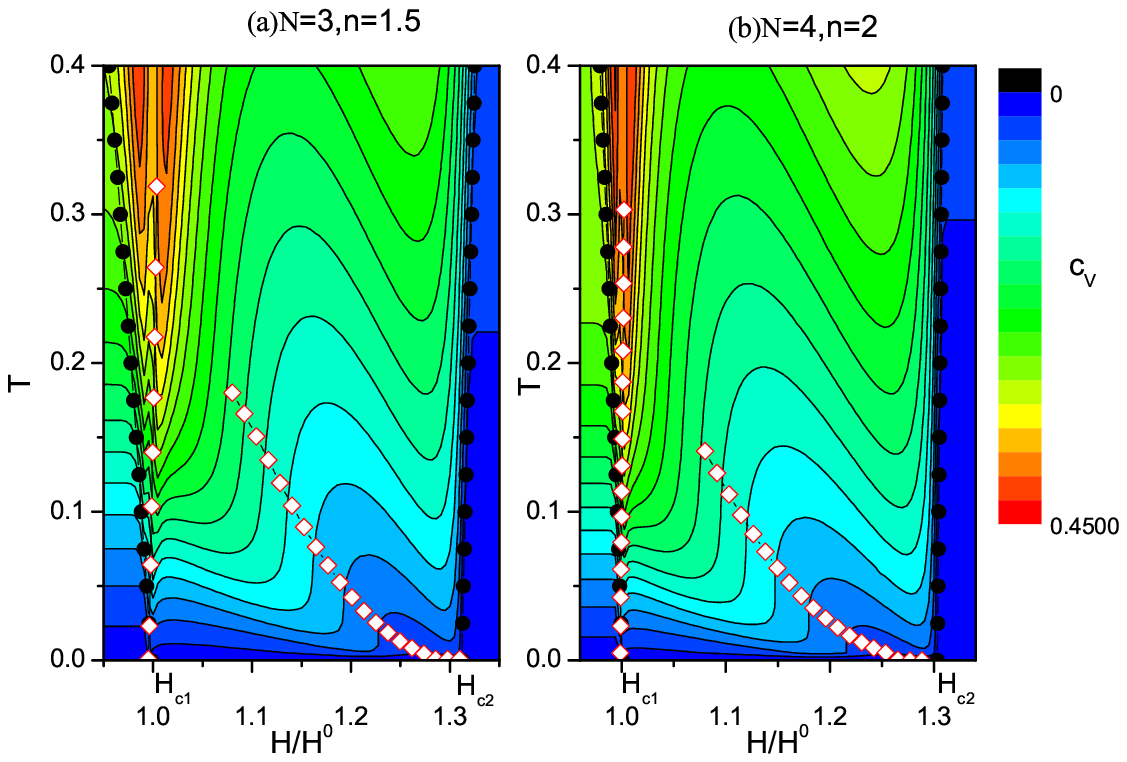}}}
\caption{Universal low temperature phase diagram: specific heat in the 
$T$-$H$ plane for $SU(3)$ and $SU(4)$. $H$ is rescaled by
$H^0=2\epsilon_{N}/(N-1)$. An asymmetric two-component TLL
remains within the regime between
$H_{c1}<H<H_{c2}$  below the white-diamond-lines. TLL of spin-neutral bound states of $N$-atoms
and TTL of fully-polarized fermionic atoms lie below the left and right 
filled black-circles, respectively.}\label{fig:cv}
\end{figure}

The analytic results (\ref{TBA}) and (\ref{FreeE}) indicate that 
the magnetization develops minima at the same
temperatures as when the two-component TLL is broken. Indeed, a
deviation from the linear temperature-dependent specific heat
$c_v=\frac{\pi
T}{3\hbar}\left(\frac{1}{v_{1}}+\frac{1}{v_{N}}\right)$ and
the magnetization minima occur around the same crossover
temperatures -- see the white-diamond-lines in Fig. \ref{fig:cv} where we
show a contour plot of the specific heat for $SU(3)$ and $SU(4)$.  The hard-core
$N$-atom composite particles and single atoms form an
asymmetric two-component TLL lying below the white-diamond-lines.
The filled-black-circles which separate the different regimes are
determined by the magnetization values $m^z=0$ and $m^z=1$.
The peaks at the critical points are expected when the
 unpaired (bound state) band starts to fill (empty) \cite{Vekua}.

To conclude, our  analytic results for universal thermodynamics and
quantum phase transitions provide a unified description of attractive 
1D $SU(N)$ fermions in the presence of external fields.
Our formulae also provide the essential equation of states for studying 
trapped $N$-component Fermi gases with population imbalance.
They thus pave the way for the study of multi-component TLL phases 
in 1D systems of ultracold fermionic atoms.

{\bf Acknowledgment.} This work has been supported by the Australian
Research Council. S. C. is supported by NSFC and 973 programs (China).
We thank M. Oshikawa, W. Vincent Liu, X.-J. Liu, H. Hu and P. Drummond for
discussions.

\clearpage

\begin{thebibliography}{99}
\bibitem{note}The details of the derivation of the TBA equations and
 the result (\ref{TBA}) will be presented elsewhere.
\end{thebibliography}
\end{document}